\documentclass[11 pt,oneside,onecolumn,a4paper]{article}
\usepackage{amsmath}
\usepackage{graphicx}
\DeclareGraphicsExtensions{.eps,.ps,.pdf}
\usepackage{bbm}
\usepackage{bm}
\usepackage{epsfig}
\usepackage[T1]{fontenc}
\usepackage{esint}
\usepackage{amssymb}

\usepackage{lipsum}

mm \leftmargin 5pt \rightmargin 5pt \hoffset= -15mm

\title{Regularities in the transformation of the oscillating decay rate in moving unstable quantum systems
}
\author{Filippo Giraldi}

\date{\small{School of Chemistry and Physics, University of KwaZulu-Natal\\ 
and National Institute for Theoretical Physics (NITheP)\\
Westville Campus, Durban 4000, South Africa}}

\begin{document}

\maketitle

\def\bbm[#1]{\mbox{\boldmath$#1$}}

\vspace{0em}




\vspace{0em}

\begin{abstract}
Decay laws of moving unstable quantum systems with oscillating decay rates are analyzed over intermediate times. The transformations of the decay laws at rest and of the intermediate times at rest, which are induced by the change of reference frame, are obtained by decomposing the modulus of the survival amplitude at rest into purely exponential and exponentially damped oscillating modes. The mass distribution density is considered to be approximately symmetric with respect to the mass of resonance. Under determined conditions, the modal decay widths at rest, $\Gamma_j$, and the modal frequencies of oscillations at rest, $\Omega_j$, reduce regularly, $\Gamma_j/\gamma$ and $\Omega_j/\gamma$, in the laboratory reference frame. Consequently, the survival probability at rest, the intermediate times at rest and, if the oscillations are periodic, the period of the oscillations at rest transform regularly in the laboratory reference frame according to the same time scaling, over a determined time window. The time scaling reproduces the relativistic dilation of times if the mass of resonance is considered to be the effective mass at rest of the moving unstable quantum system with relativistic Lorentz factor $\gamma$.
\end{abstract}

\maketitle

\section{Introduction}\label{1}
\vspace{-0em}

In recent years the experimental work which has been devoted to analyze the decays of unstable systems has shown various oscillating behaviors of the decay laws \cite{GSI1,GSI2,OscExperiments1}. Oscillations of the decay probability density, or, equivalently, the decay rate, have been detected over short times in the electron capture decays of Hydrogen-like ions \cite{GSI1,GSI2}. These short-time oscillations are superimposed on the canonical exponential decay law and are known as the GSI anomaly. Refer to \cite{GSI1,GSI2} for details. Over much longer time scale, small periodical deviations are observed in the exponential decay of the unstable nuclei ${}^{32} {\rm Si}$ \cite{OscExperiments1}. The period of the superimposed oscillations is one year, while the half-life of the unstable state is approximately $170$ years. Refer to \cite{OscExperiments1} for details. Oscillating decay laws of unstable quantum systems are obtained in Refs. \cite{GPoscillatingDecaysQM2012,GPoscillatingDecaysQM2012PoS} by introducing deviations from the Breit-Wigner mass (energy) distribution density. These deviations are proposed as an explanation of the GSI anomaly. See Refs. \cite{GPoscillatingDecaysQM2012,GPoscillatingDecaysQM2012PoS} for details. 

Decay processes are often detected in laboratory reference frames where the unstable systems are moving at relativistic or ultrarelativistic velocities. For example, the relativistic Lorentz factor of the Hydrogen-like ions which are produced in the GSI experiment is equal to $1.43$. See Table 2 of Ref. \cite{GSI1} for details. Due to the change of reference frame, the decay laws at rest are detected as transformed in the laboratory reference frame. This transformation is described via quantum theory and special relativity in case the unstable system moves with constant linear momentum in the laboratory reference frame \cite{Khalfin,BakamjianPR1961RQT,ExnerPRD1983,HEP_Stef1996,HEP_Shir2004,HEP_Shir2006,TD_StefanovichXivHep2006,UrbPLB2014,GiacosaAPPB2016,GiacosaAPPB2017,UrbAPB2017,GiacosaAHEP2018}. The survival amplitude at rest transforms in the laboratory reference frame according to an integral form which involves the MDD and depends on the linear momentum of the moving unstable quantum system.

The appearance of the relativistic dilation of times in the decay laws of moving unstable particles remains a matter of central interest. See Refs. \cite{HEP_Stef1996,HEP_Shir2004,TD_MuonsNat,HEP_Shir2006,HEP_Shir2009,FlemingXiv2011,UrbPLB2014,GiacosaAPPB2016,GiacosaAPPB2017,UrbAPB2017,GiacosaAHEP2018}, to name but a few. Recent analysis has shown that the long-time inverse-power-law decays at rest of moving unstable quantum systems transform in the laboratory reference frame, approximately, according to a scaling relation \cite{Gxiv2018}. The scaling factor is determined by the (non-vanishing) lower bound of the mass spectrum and by the linear momentum, and consists in the ratio of the asymptotic value of the instantaneous mass and of the instantaneous mass at rest of the moving unstable quantum system \cite{UrbPLB2014,UrbAPB2017,UrbanowskiEPJD2009,UrbanowskiCEJP2009,UrbanowskiPRA1994,GEPJD2015}. The scaling relation reproduces, approximately, the relativistic dilation of times if the (non-vanishing) lower bound of the mass spectrum is considered to be the effective mass at rest of the moving unstable quantum system. See Ref. \cite{Gxiv2018} for details. Oscillating behaviors of the survival probability appear if the unstable quantum system is prepared in the superposition of two eigenstates of the Hamiltonian with different eigenvalues \cite{HEP_Shir2006,ShirokovNaumov2006}. In the laboratory reference frame the transformed survival probability shows a dilation of times which, in general, is not Einsteinian. See Refs. \cite{HEP_Shir2006,ShirokovNaumov2006} for details. The relativistic dilatation of times is found, approximately, if the initial superposition consists in two, approximately orthogonal, unstable quantum states with Breit-Wigner forms of MDDs and the two masses of resonance are approximately equal \cite{HEP_Shir2004}. If the two MDDs are bounded from below and exhibit thresholds, the survival probability is described in the laboratory reference frame by damped oscillations, in case the lower bounds of the two mass spectra differ \cite{GJPA2018}. The frequency of the damped oscillations diminishes if it is compared to the frequency of the oscillations at rest. The transformed frequency tends to vanish in the ultrarelativistic limit. The lower bounds of the mass spectra and the linear momentum of the moving unstable quantum state determine the ratio of the two frequencies.

Recently, the decay laws of moving unstable quantum systems have been studied over intermediate times by decomposing the modulus of the survival amplitude at rest into superpositions of exponential modes via the Prony analysis \cite{Gxiv2019}. The transformation of the intermediate times is described by expressing the survival probability $\mathcal{P}_p(t)$, which is detected in the laboratory reference frame $\mathfrak{S}_p$, where the unstable system moves with constant linear momentum $p$, as the transformed form $\mathcal{P}_0\left(\varphi_p(t)\right)$ of the survival probability at rest $\mathcal{P}_0(t)$, which is detected in the rest reference frame $\mathfrak{S}_0$. Over a determined time window, this function grows linearly. The corresponding scaling law reproduces the relativistic dilation of times, approximately, over the time window if the mass of resonance of the MDD is interpreted as the effective mass at rest of the moving unstable quantum system. Refer to \cite{Gxiv2019} for details.

As a continuation of the above-described scenario, here, we consider moving unstable quantum systems which exhibit in the rest reference frame $\mathfrak{S}_0$ oscillating decay rates. We intend to evaluate the transformed decay laws and the transformation of times in case the decay laws at rest are decomposed into superpositions of purely exponential modes and exponentially damped oscillating modes. We aim to find the conditions under which the relativistic dilation of times appears in the transformation of times.

The paper is organized as follows. Section \ref{2} is devoted to the decay laws of moving unstable quantum systems and to the general transformation which is due to the change of reference frame. In Section \ref{3}, the transformation of the decay laws with oscillating decay rates is determined via the transformed exponential and exponentially damped oscillating modes. Section \ref{4} is devoted to the appearance of the relativistic time dilation in the genera transformation of times. Summary and conclusions are reported in Section \ref{5}. Demonstrations of the results are provided in appendix.

\section{Moving unstable quantum systems and oscillating decay rate}\label{2}

For the sake of clarity, we report below some details about the general transformation of the decay laws at rest which provides the decay laws in the laboratory reference frame $\mathfrak{S}_p$. The description is performed by following Ref. \cite{UrbAPB2017}. Let the state kets $|m,p\rangle$ belong to the Hilbert space $\mathcal{H}$ of the quantum states of the unstable system and be the common eigenstates of the linear momentum $P$ and of the Hamiltonian $H$ self-adjoint operator. The corresponding eigenvalues are respectively $p$, i.e., $P|m,p\rangle =p |m,p\rangle$, and $E(m,p)$, i.e., $H|m,p\rangle =E(m,p) |m,p\rangle$, for every value $m$ which belongs to the continuous spectrum of the Hamiltonian. Let $|\phi\rangle$ be the initial state of the unstable quantum system. This state ket belongs to the Hilbert space $\mathcal{H}$ and is represented in terms of the eigenstates $|m,0\rangle$ of the Hamiltonian, $|\phi\rangle=\int_{\mu_0}^{\infty} \langle 0,m||\phi\rangle |m,0\rangle dm$, where $\langle 0,m|$ is the bra of the state ket $|m,0\rangle$.

In the rest reference frame $\mathfrak{S}_0$ of the moving unstable quantum system, the survival amplitude at rest $A_0(t)$ is given by the following form, $A_0(t)=\langle \phi| e^{-\imath H t} |\phi\rangle$, where $\imath$ is the imaginary unit. The completeness of the eigenstates of the Hamiltonian leads to the following integral expression of the survival amplitude at rest \cite{UrbAPB2017,UrbPLB2014,UrbanowskiEPJD2009,UrbanowskiCEJP2009,FondaGirardiRiminiRPP1978},
\begin{eqnarray}
A_0(t)=\int_{\rm{spec}\left\{H\right\} }\omega\left(m\right) 
\exp\left(-\imath m t\right) dm. \label{A0Int}
\end{eqnarray}
The domain of integrations, $\rm{spec}\left\{H\right\}$, is the  spectrum of the Hamiltonian $H$, which is assumed to be continuous. The function $\omega\left(m\right)$ represents the MDD of the unstable quantum system, $\omega\left(m\right)=\left|\langle 0,m||\phi\rangle\right|^2$. The initial state and the Hamiltonian of the system determines the MDD via the eigenstates $|m,0\rangle$. In the rest reference frame $\mathfrak{S}_0$ of the moving unstable quantum system, the non-decay or survival probability is given by the square modulus of the survival amplitude at rest, $\mathcal{P}_0(t)=\left|A_0(t)\right|^2$, and is referred to as survival probability at rest. Expression (\ref{A0Int}) of the survival amplitude provides oscillations of the decay rate, which are superimposed to the canonical exponential decay law, in case the Breit-Wigner form of the MDD is cut on the left and right side of the peak \cite{GPoscillatingDecaysQM2012,GPoscillatingDecaysQM2012PoS}. The high-mass cutoff determines the oscillatory behavior of the decay rate and does not alter significantly the exponential decay, while the low-mass cutoff determines the transition from the exponential to the inverse-power-law regime of the decay. The MDD is symmetric with respect to the mass of resonance, in general, in case the mass-dependent cutoff of the Breit-Wigner form exhibits the same symmetry. In particular, the Breit-Wigner MDD is symmetric if it is cut by the smooth Fermi-like cutoff function. See Refs. \cite{GPoscillatingDecaysQM2012,GPoscillatingDecaysQM2012PoS} for details.

In the laboratory reference frame $\mathfrak{S}_p$ the unstable system is described by the state ket $|\phi_p\rangle$ which is an eigenstate of the linear momentum $P$ with eigenvalue $p$. The transformed survival amplitude $A_p(t)$ is given by the following expression, $\langle \phi_p|e^{-\imath H t}|\phi_p\rangle$, and is represented by the integral form below,
\begin{eqnarray}
A_p(t)=\int_{\rm{spec}\left\{H\right\} } \omega\left(m \right)
\exp\left(-\imath \sqrt{ p^2+m^2}t\right) d m.
\label{Aptdef}
\end{eqnarray}
The survival probability, $\mathcal{P}_p(t)$, is provided by the square modulus of the above expression, $\mathcal{P}_p(t)=\left|A_p(t)\right|^2$. The integral expression (\ref{Aptdef}) of the survival probability has been obtained in various ways by adopting quantum theory and special relativity. See 
Refs. \cite{HEP_Stef1996,HEP_Shir2004,HEP_Shir2006,GiacosaAPPB2016,GiacosaAPPB2017,UrbAPB2017,GiacosaAHEP2018} for details.

An analytical form of the transformed survival amplitude $A_p(t)$ is obtained in Ref. \cite{HEP_Shir2004} for the truncated form of the Breit-Wigner MDD  which vanishes for negative values of the mass spectrum. Over sufficiently long times the survival amplitude results in a superposition of a dominant exponential decay and a dominant inverse power law. 
If the decay width is very small compared to the mass of resonance the exponential term of the survival amplitude dominates over the inverse power law for various lifetimes \cite{HEP_Shir2004,FondaGirardiRiminiRPP1978} and the survival probability exhibits an approximate exponential decay over these times. Under determined conditions, the relativistic dilation of times is approximately recovered. See Ref. \cite{HEP_Shir2004} for details. 

In Ref. \cite{Gxiv2019} the transformed decay laws are described analytically over intermediate times by approximating the modulus of the survival amplitude at rest with superpositions of exponential modes via the Prony analysis. This description relies upon the following assumptions. Over intermediate times, the decay laws are mainly determined by the behavior of the MDD around the mass of resonance $M$ and the contribution of the unphysical negative values of the mass spectrum to the decay laws is negligible. The MDD is considered to be approximately symmetric with respect to the mass of resonance, 
\begin{eqnarray}
\omega \left(M+m^{\prime}\right)=\omega \left(M-m^{\prime}\right). \label{symmMDD}
\end{eqnarray}
These assumptions are founded on the following observations \cite{MDDlorentzianGislasonPRA1991,NnExpKumarAJP1992,MDDlorentzianJacobovitsAJP1995,GPoscillatingDecaysQM2012,GiacosaAPPB2016,GiacosaAPPB2017}. Purely exponential decays of the survival probability are obtained from Lorentzian MDDs through the complex-valued and simple pole of the Lorentzian function. Lorentzian MDDs are symmetric with respect to the mass of resonance. The support of these MDDs is the whole real line and the contribution of the unphysical negative values of the mass spectrum to the survival amplitude is negligible. This condition is satisfied in case the decay widths of the exponential modes are small compared to the mass of resonance. This approach provides an integral form of the MDD in term of the modulus of the survival amplitude at rest,
\begin{eqnarray}
\omega\left(M+m^{\prime}\right)=\frac{1}{\pi}\left|
\int_0^{\infty}\sqrt{\mathcal{P}_0\left(t^{\prime}\right)}
 \cos\left(m^{\prime}t^{\prime}\right) d t^{\prime}\right|
. \label{MMDP0int1}
\end{eqnarray}
In this way, the transformed survival probability is related to the modulus of the survival amplitude at rest by the integral form below,
\begin{eqnarray}
\hspace{-0em}\mathcal{P}_p(t)=\left|\frac{2}{\pi}\int_0^{\infty}
\exp\left(-\imath \sqrt{p^2+m^2}t\right)
dm \int_0^{\infty}\sqrt{\mathcal{P}_0\left(t^{\prime}\right)} \cos\left(M t^{\prime}\right)\cos\left(m t^{\prime}\right) dt^{\prime} \right|^2. \label{PpP0Int1}
\end{eqnarray}
In the rest reference frame $\mathfrak{S}_0$, the exponential-like regime represents the condition over which the decay is purely exponential, or slower, but has not yet become the inverse power law. Relying on the above-reported assumptions, the modulus of the survival amplitude at rest is approximated by a superposition of a finite number of exponential modes over the intermediate times.
 The above approximation is provided by the Prony analysis of the modulus of the survival amplitude at rest. The transformed expression of the survival probability is evaluated via Eq. (\ref{PpP0Int1}) and from the analysis which is performed in Ref. \cite{HEP_Shir2004}. In the laboratory reference frame $\mathfrak{S}_p$, the survival probability $\mathcal{P}_p(t)$ is described, under determined conditions, by a superposition of exponential modes over an estimated time window. The survival probability $\mathcal{P}_p(t)$ is approximately related to the survival probability at rest $\mathcal{P}_0 \left(t\right)$ by the scaling relation which reproduces, approximately, the relativistic dilation of times, if the mass of resonance $M$ of the MDD is interpreted as the effective mass at rest of the unstable quantum systems which moves with constant linear momentum $p$. See Ref. \cite{Gxiv2019} for details.

In the following, we intend to use Eq. (\ref{PpP0Int1}) in order to evaluate the survival probability $\mathcal{P}_p(t)$, which is detected in the laboratory reference frame $\mathfrak{S}_p$ over intermediate times, by decomposing the modulus of the survival amplitude at rest $\sqrt{\mathcal{P}_0(t)}$ into purely exponential and exponentially damped oscillating modes. It is assumed that the MDD is approximately symmetric with respect to the mass of resonance, Eq. (\ref{symmMDD}). This assumption relies on the observations which are reported at the end of the second paragraph and at the beginning of the fifth paragraph of the present Section.

\section{Transformation of oscillating decay laws}\label{3}

At this stage, we start our analysis by considering decay laws at rest with oscillating decay rates. For the sake of shortness, here, these decay laws are referred to as oscillating. We intend to study how this kind of decays transforms by changing the 
reference frame. We start from the simplest form of oscillating decay law at rest which is obtained by adding an exponentially damped oscillating term to the exponentially decaying modulus of the survival amplitude,
\begin{eqnarray}
\sqrt{\mathcal{P}_0(t)}=\exp\left(-\frac{\Gamma}{2} t\right)\left(1-a + a \cos \left(\Omega t\right)\right).
\label{sqrtP0toy1}
\end{eqnarray}
This form is generalized in the next Section by approximating the modulus of the survival amplitude  at rest via the superposition of an arbitrary finite number of purely exponential modes and exponentially damped oscillating modes. The restrictions $1/2>a>0$, $\Omega<M$, $\Gamma/\left(M-\Omega\right) \ll 1$ and $\Gamma>2 a \Omega/\sqrt{1-2a}$ are required for the parameter $a$, for the frequency $\Omega$ of the oscillations and for the decay width $\Gamma$. In this way, the survival probability at rest $\mathcal{P}_0(t)$ fulfills the canonical proprieties. The function $\mathcal{P}_0(t)$ is positive, decreases monotonically, $\dot{\mathcal{P}}_0(t)<0$ for every $t>0$, and takes the value $\mathcal{P}_0(0)=1$, in addition to $\mathcal{P}_0\left( \infty\right)=0$. The survival probability at rest $\mathcal{P}_0(t)$ is the sum of the purely exponential term $\mathcal{P}^{\rm (exp)}_0(t)$ and of the exponentially damped oscillating term $\mathcal{P}^{\rm (osc)}_0(t)$,
\begin{eqnarray}
&&\mathcal{P}_0(t)= \mathcal{P}^{\rm (exp)}_0(t)+ \mathcal{P}^{\rm (osc)}_0(t),
\label{P0ExpOscapprox}
\end{eqnarray}
where
\begin{eqnarray}
&&\mathcal{P}^{\rm (exp)}_0(t)= \left(\left(1-a\right)^2+\frac{a^2}{2}\right)
\exp \left(-\Gamma t\right),
 \label{P0Expapprox}\\
&&\mathcal{P}^{\rm (osc)}_0(t)= \exp \left(-\Gamma t\right)\left(2 a \left(1-a\right)
\cos \left(\Omega t\right)+ \frac{a^2}{2}\cos \left(2\Omega t\right)\right).
\label{P0Oscapprox}
\end{eqnarray}
The decay probability density at rest, or the decay rate at rest, is defined as the opposite of the derivative of the survival probability at rest and is provided by the oscillating form below,
\begin{eqnarray}
\dot{\mathcal{P}}_0(t)=-\exp\left(-\Gamma t\right)
\left(1-a+a 
\cos \left(\Omega t\right)\right)
\left(\lambda_{1}+\lambda_{2} 
\cos \left(\Omega t-\beta\right)\right). \label{D1P0toy1}
\end{eqnarray}
The coefficients $\lambda_{1}$ and 
$\lambda_{2}$, and the angle $\beta$ read
\begin{eqnarray}
&&\hspace{-4em}\lambda_{1}=
\Gamma \left(1-a\right),\hspace{1em}\lambda_{2}=a\sqrt{\Gamma^2+4\Omega^2},\hspace{1em} \beta = \arccos \frac{\Gamma}{\sqrt{\Gamma^2+4 \Omega^2}}.\nonumber
\end{eqnarray}
Again, we stress that Eq. (\ref{sqrtP0toy1}) can not approximate the survival probability over very short and very long times. In fact, Eq. (\ref{D1P0toy1}) provides a negative value of the decay rate, $\dot{\mathcal{P}}_0(0)=-\Gamma$, instead of vanishing, for $t=0$.

In the laboratory reference frame $\mathfrak{S}_p$ the survival probability $\mathcal{P}_p(t)$ is approximated by the expression below,
\begin{eqnarray}
&&\hspace{-2em}\mathcal{P}_p(t)\simeq\Bigg|
 K\left(M,\Gamma,p,\Omega,a,t\right)
+\imath \frac{p \Gamma}{\pi M^2} \Phi\left(M,p,\Omega,a,t\right)\Bigg|^2, \label{PptHYJosc}
\end{eqnarray}
over times $t$ such that either $t>1/\left(10 \Gamma\right)$ or $\left(M-\Omega\right) t \gg 1$. The function $K\left(M,\Gamma,p,\Omega,a,t\right)$ and $\Phi\left(M,p,\Omega,a,t\right)$ are defined as below,
\begin{eqnarray}
&&\hspace{-1.0em}
 K\left(M,\Gamma,p,\Omega,a,t\right)=
\left(1-a\right)\exp\left(- \frac{1}{2}\Upsilon\left(M,\Gamma,p\right)t\right)+\frac{a}{2}\Bigg(
\exp\left(- \frac{1}{2}\Upsilon\left(M-\Omega,\Gamma,p\right)t\right)\nonumber \\&&\hspace{8.4em}+\exp\left( -\frac{1}{2}\Upsilon\left(M+\Omega,\Gamma,p\right)t\right)\Bigg),
\label{Pi}\\
&&\hspace{-1.0em}
 \Phi\left(M,p,\Omega,a,t\right)=\left(1-a\right)\Xi\left(M,p,t\right)+\frac{a}{2}\Bigg(\frac{\Xi\left(M-\Omega,p,t\right)}{\left(1-\Omega/M\right)^2}+\frac{\Xi\left(M+\Omega,p,t\right)}{\left(1+\Omega/M\right)^2}\Bigg).\label{Phi}
\end{eqnarray}
The function $\Upsilon\left(M,\Gamma,p\right)$ is defined for every value of the mass of resonance $M$, of the decay width $\Gamma$ and of the linear momentum $p$, 
by the following expression \cite{Gxiv2019},
\begin{eqnarray}
\Upsilon\left(M,\Gamma,p\right)=\Lambda_-\left(M,\Gamma,p\right)
 + \imath \Lambda_+\left(M,\Gamma,p\right), \label{Upsilon}
\end{eqnarray}
where 
\begin{eqnarray}
&&\hspace{-3em}\Lambda_{\mp}\left(M,\Gamma,p\right)=\sqrt{2\left(\sqrt{\left(M^2-\frac{\Gamma^2}{4}+p^2\right)^2+M^2 \Gamma^2}\mp\left(M^2-\frac{\Gamma^2}{4}+p^2\right)\right)}. \label{Lambdamp}
\end{eqnarray}
The function $\Xi\left(M,p,t\right)$ is defined, for every value of the mass of resonance $M$, of the decay width $\Gamma$ and of the linear momentum $p$, via the Bessel Function $J_1\left(pt\right)$, the modified Bessel function $Y_1\left(pt\right)$
and the Struve function $\mathbf{H}_1\left(pt\right)$ as below \cite{NISThandbook,GradRyzhandBook,AbrSteg},
\begin{eqnarray}
&&\hspace{-2.2em}\Xi\left(M,p,t\right)=\frac{\pi }{2}\left(\mathbf{H}_1\left(pt\right)-\imath J_1\left(pt\right)\right)-1+
\frac{1-p^2/M^2}{\left(1+p^2/M^2\right)^2} \nonumber \\ &&\hspace{3.6em}\times \Big(1+\frac{\pi }{2}\left(Y_1\left(pt\right)-\mathbf{H}_1\left(pt\right)\right)\Big). 
\label{Xi}
\end{eqnarray}
Due to the oscillations, additional terms appear in the expression of the transformed survival probability. These terms are related to the shifted masses of resonance $M^-$ and $M^+$ which are determined by the frequency $\Omega$ of the oscillations, 
$$M^{\mp}=M \mp \Omega.$$ We remind that the condition $\Omega<M$ is required.

\subsection{Time window for exponentially damped oscillating decays in the laboratory reference frame }\label{31}

At this stage, we search for times over which the oscillating decay laws are exponentially damped in the laboratory reference frame $\mathfrak{S}_p$. Here, we refer to these times, if they exist, as the exponential times of the transformed oscillating decays. For the sake of convenience, we introduce the shifted relativistic Lorentz factors $\gamma^+$ 
and $\gamma^-$ and the shifted decay widths $\Gamma^+$ and $\Gamma^-$, 
\begin{eqnarray}
\gamma^{\mp}=\sqrt{1+\frac{p^2}{\left(M^{\mp}\right)^2}}, \hspace{1em}\Gamma^{\mp}= \frac{\gamma}{\gamma^{\mp}} \Gamma. \nonumber
\end{eqnarray}
The function $\Pi\left(M,\Gamma,p,\Omega,a,t\right)$, which appears in Eq. (\ref{PptHYJosc}), provides purely exponential decays with decay widths
 $\Gamma^{-}/\gamma$, $\Gamma/\gamma$, $\Gamma^{+}/\gamma$, and exponentially damped oscillations with frequencies $\left|M^- \gamma^--M \gamma\right|$, $\left|M^+ \gamma^+-M \gamma\right|$, $\left|M^- \gamma^--M^+ \gamma^+\right|$, due to the condition $\Gamma/M^- \ll 1$.

In case the frequency of the oscillations at rest is small compared to the mass of resonance, $\Omega \ll M$, and if the parameter $a$ 
is not too close to the value $1/2$, the modulus of the transformed survival amplitude is approximated via the simple form below,
\begin{eqnarray}
\mathcal{P}_p(t)\simeq 
\exp\left(- \frac{\Gamma}{ \gamma}t\right) \left(1-a+a \cos\left(\frac{\Omega}{\gamma} t\right)\right)^2,
\label{sqrtPptoy1}
\end{eqnarray}
over the time window 
\begin{eqnarray}\frac{2 \zeta_{\rm max} }{\Gamma}\gamma \gtrsim t\gtrsim\frac{2 \zeta_{\rm min}}{\Gamma}\gamma, \label{ExpTExp0}
\end{eqnarray}
on condition that the following constraints are fulfilled, $M^- t \gg 1$ or 
$t >1/\left(10 \Gamma\right)$, and $p t \gg 1$.
The allowed values of the parameter $a$ are estimated by studying the order of magnitude of the parameter $\xi^{\prime}$ which is defined as below,
\begin{eqnarray}
\xi^{\prime}=\frac{\Gamma W\left(M,\Omega,a\right)}{2M\left(1-2a\right)}\sqrt{\frac{\Gamma }{\pi  M}\sqrt{1-\frac{1}
{\gamma^2}}}. \label{xiprime}
\end{eqnarray} The function $W\left(M,\Omega,a\right)$ is defined in appendix.
If $\xi^{\prime} \ll 10^{-2}$ and $\Omega \ll M$, the transformed survival probability is given by Eq. (\ref{sqrtPptoy1}) over the times $t$ of the exponential time window (\ref{ExpTExp0}) such that either $t >1/\left(10 \Gamma\right)$ or $M^- t \gg 1$, and $p t \gg 1$. This technique is adopted in Ref. \cite{Gxiv2019} in order to estimate the set of times over which the decay laws consist, approximately, in superpositions of exponential modes in the laboratory reference frame $\mathfrak{S}_p$. For example, let the order of magnitude of the ratio $\Gamma/M$ be equal to or less than $\left(-3\right)$ and let the parameter $a$ fulfill the constraint $0<a<\left(1-10^{-1}\right)/2$. The condition $t>1/\left(10 \Gamma_1\right)$ is fulfilled over the time window (\ref{ExpTExp0}) if $20 \zeta_{\rm min}\gamma>1$. If this condition is not realized, the constraint $M^- t \gg 1$ is fulfilled over the time window if $2 \zeta_{\rm min}\gamma\gg\Gamma/M_-$. The condition 
$p t \gg 1$ holds over the time window if $2 \zeta_{\rm min}\gamma \sqrt{\gamma^2-1}\gg\Gamma/M$. The nonrelativistic limit, $\gamma\to 1^+$, is excluded. Usually, the above constraint hold except for strong decays and nonrelativistic regime, $\gamma \simeq 1$ or, equivalently, $p \ll M$. If these constraints do not hold, a different order of magnitude of the parameter $\xi^{\prime}$ is chosen and the values of the parameters $\zeta_{\rm min}$ and $\zeta_{\rm max}$ must be changed accordingly. Refer to \cite{Gxiv2019} for details.

In case $\Omega \ll M$ and if the parameter $a$ is not too close to the value $1/2$ in such a way that $\xi^{\prime} \ll 10^{-2}$, the analysis which is performed in the previous paragraph suggests that the survival probability at rest transforms according to the following scaling relation, 
\begin{eqnarray}
\mathcal{P}_p(t)\simeq \mathcal{P}_0 \left(\frac{t}{\gamma}\right), \label{PpPprel}
\end{eqnarray}
 over the times $t$ of the window (\ref{ExpTExp0}) such that either $M^- t \gg 1$ or 
$t >1/\left(10 \Gamma\right)$, and $p t \gg 1$. Over the selected times both the exponential-like term $\mathcal{P}^{\rm (exp)}_p(t)$ and the damped oscillating term $\mathcal{P}^{\rm (osc)}_p(t)$ transform, independently and regularly, according to the same scaling relation, 
\begin{eqnarray}
&&\mathcal{P}_p(t)\simeq \mathcal{P}^{\rm (exp)}_p(t)+ \mathcal{P}^{\rm (osc)}_p(t),
\label{PPExpOscapprox}
\end{eqnarray}
where
\begin{eqnarray}
&&\mathcal{P}^{\rm (exp)}_p(t)\simeq \mathcal{P}^{\rm (exp)}_0 \left(\frac{t}{\gamma}\right),
 \label{PpExpapprox}\\
&&\mathcal{P}^{\rm (osc)}_p(t)\simeq \mathcal{P}^{\rm (osc)}_0 \left(\frac{t}{\gamma}\right).
\label{PpOscapprox}
\end{eqnarray}
The decay width $\Gamma$ and the frequency $\Omega$ of the oscillating decay law at rest (\ref{sqrtP0toy1}) transform regularly, by changing reference frame, in the values $\Gamma/\gamma$ and $\Omega/\gamma$, respectively. The transformed decay width, $\Gamma/\gamma$, and the transformed frequency of the oscillation, $\Omega/\gamma$, tend to vanish in the ultrarelativistic limit, $p \gg M$. Approximately, the transformed period $\mathcal{T}_p$ of the oscillations depends regularly on the period of oscillation $\mathcal{T}_0$ at rest,
\begin{eqnarray}
\mathcal{T}_p\simeq\gamma \mathcal{T}_0, \label{TpT0}
\end{eqnarray}
where $\mathcal{T}_0=2 \pi/\Omega$. Obviously, the transformed period diverges in the ultrarelativistic limit, $p \gg M$.

\begin{figure*}
 \includegraphics[width=0.5\textwidth]{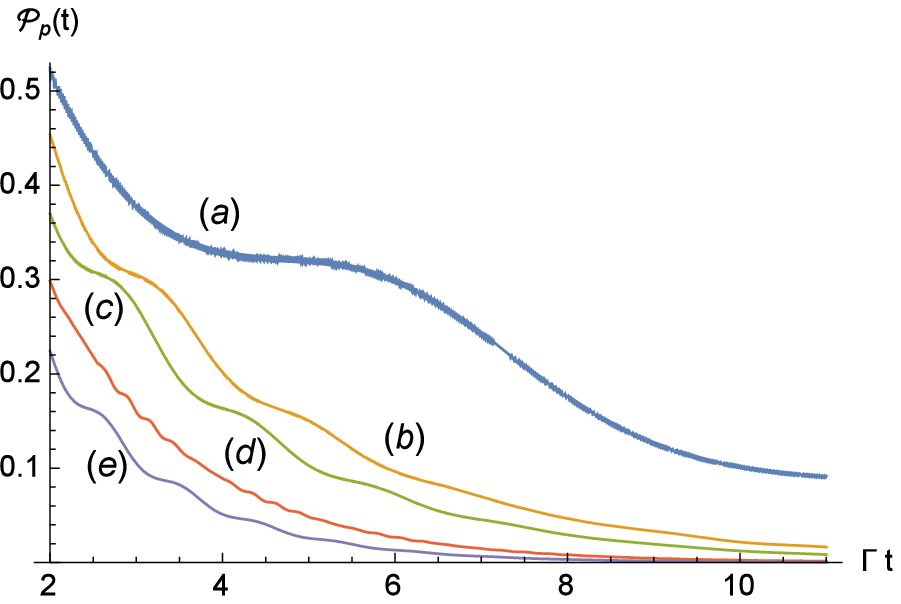}
\caption{(Color online) Transformed survival probability $\mathcal{P}_p(t)$ versus $ \Gamma t$, for $ 2 \leq \Gamma t \leq 11$, and different values of the ratios $p/\Gamma$ and $M/\Gamma$ and of the corresponding Lorentz factor $\gamma$. The survival probability at rest $\mathcal{P}_0(t)$ is described via Eq. (\ref{sqrtP0toy1}) with different values of the frequency $\Omega$ and of the amplitude $a$. Curve $(a)$ corresponds to $p/ \Gamma=150$, $M/\Gamma=30$, $\gamma\simeq 5.0990$, $\Omega/\Gamma=5$ and $a=0.09$. Curve $(b)$ corresponds to $p/ \Gamma=200$, $M/\Gamma=80$, $\gamma\simeq 2.6926$, $\Omega/\Gamma=10$ and $a=0.04$. Curve $(c)$ corresponds to $p/ \Gamma=210$, $M/\Gamma=100$, $\gamma\simeq 2.3259$, $\Omega/\Gamma=10$ and $a=0.04$. Curve $(d)$ corresponds to $p/ \Gamma=200$, $M/\Gamma=150$, $\gamma\simeq 1.6667$, $\Omega/\Gamma=40$ and $a=0.01$. Curve $(e)$ corresponds to $p/ \Gamma=100$, $M/\Gamma=100$, $\gamma0\sqrt{2}$, $\Omega/\Gamma=10$ and $a=0.04$.}
\label{fig1}
\end{figure*}

\begin{figure*}
 \includegraphics[width=0.5\textwidth]{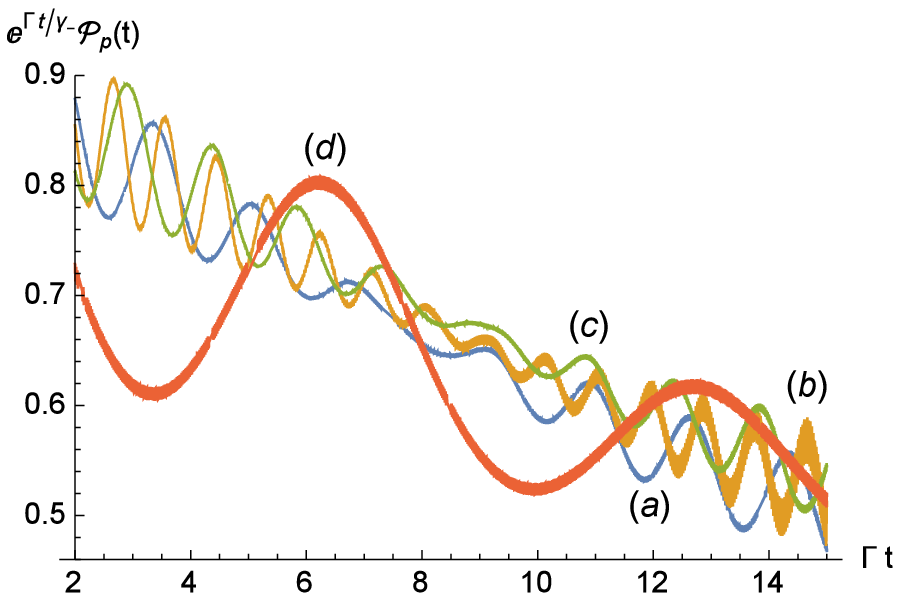}
\caption{(Color online) Quantity $\exp\left(\Gamma t /\gamma_-\right)\mathcal{P}_p(t)$ versus $ \Gamma t$, for $ 2 \leq \Gamma t \leq 15$, and different values of the ratios $p/\Gamma$ and $M/\Gamma$ and of the corresponding Lorentz factor $\gamma$. The survival probability at rest $\mathcal{P}_0(t)$ is described via Eq. (\ref{sqrtP0toy1}) for different values of the frequency $\Omega$ and of the amplitude $a$. Curve $(a)$ corresponds to $p/ \Gamma=200$, $M/\Gamma=80$, $\gamma\simeq 2.6926$, $\Omega/\Gamma=10$ and $a=0.04$. Curve $(b)$ corresponds to $p/ \Gamma=100$, $M/\Gamma=100$, $\gamma=\sqrt{2}$, $\Omega/\Gamma=10$ and $a=0.04$. Curve $(c)$ corresponds to $p/ \Gamma=210$, $M/\Gamma=100$, $\gamma\simeq 2.3259$, $\Omega/\Gamma=10$ and $a=0.04$. Curve $(d)$ corresponds to $p/ \Gamma=150$, $M/\Gamma=30$, $\gamma\simeq 5.0990$, $\Omega/\Gamma=5$ and $a=0.09$.}
\label{fig2}
\end{figure*}

Numerical analysis of the survival probability $\mathcal{P}_p(t)$ is performed in Figure \ref{fig1} in case the survival probability at rest $\mathcal{P}_0(t)$ is obtained from Eq. (\ref{sqrtP0toy1}). The damped oscillating behavior which is displayed in Figure \ref{fig2} is in accordance with the asymptotic analysis which is described by Eqs. (\ref{sqrtPptoy1})-(\ref{TpT0}). The transformed periods of the oscillations agree with Eq. (\ref{TpT0}). 

 \subsection{Transformation of times and relativistic time dilation in the oscillating decay laws}\label{32}

At this stage, we study the transformation of times in case the oscillating decay law at rest is obtained from Eq. (\ref{sqrtP0toy1}). This transformation is described by the function $\varphi_p(t)$, which is introduced in Ref. \cite{Gxiv2019} via the following relation,
\begin{eqnarray}
\mathcal{P}_p(t)=\mathcal{P}_0\left(\varphi_p(t)\right). \label{PpP0def}
\end{eqnarray}
The survival probability at rest $\mathcal{P}_0(t)$ is an invertible function of time since it is canonically decreasing. Therefore, the function $\varphi_p(t)$ is properly defined by the following expression,
\begin{eqnarray}
\varphi_p(t)=\mathcal{P}^{-1}_0\left(\mathcal{P}_p(t)\right), \label{phipP0Pp}
\end{eqnarray}
for every $t \geq 0$. In general, for the oscillating decay law (\ref{sqrtP0toy1}) the inverse function $\mathcal{P}^{-1}_0\left(r\right)$ is computed numerically.
Once the function $\mathcal{P}^{-1}_0\left(r\right)$ is obtained, the function $\varphi_p(t)$ is evaluated under the conditions $\Omega<M$, $\Gamma/M^- \ll 1$, and either $M^- t \gg 1$ or $t>1/\left(10 \Gamma\right)$, via Eqs. (\ref{PptHYJosc})-(\ref{Phi}), 
\begin{eqnarray}
&& \hspace{-2em}\varphi_p(t)\simeq \mathcal{P}^{-1}_0\Bigg(\Bigg|
 K\left(M,\Gamma,p,\Omega,a,t\right)
+\imath \frac{p \Gamma}{\pi M^2} \Phi\left(M,p,\Omega,a,t\right)\Bigg|^2\Bigg). \label{phipP0PpHYJosc1}
\end{eqnarray}

In case the frequency $\Omega$ of the oscillations at rest is small compared to the mass of resonance, $\Omega \ll M$, and if $\xi^{\prime} \ll 10^{-2}$, the survival probability transforms according to the scaling relation (\ref{PpPprel}) over the times $t$ which belong to the time window (\ref{ExpTExp0}) and such that either $M^- t \gg 1$ or $t >1/\left(10 \Gamma\right)$, and $p t \gg 1$. The nonrelativistic regime, $p\ll M$, is excluded. According to the scaling relation (\ref{PpPprel}), the function $\varphi_p(t)$ is approximately linear over the exponential times,
\begin{eqnarray}
\varphi_p(t) \simeq \frac{t}{\gamma}. \label{varphilinear}
\end{eqnarray}
If the mass of resonance $M$ is considered to be the effective mass at rest of the unstable quantum systems which moves with constant linear momentum $p$, the scaling relation (\ref{PpPprel}) reproduces, approximately, the relativistic dilation of times. Over these times, the period of the oscillations is the relativistic dilation of the period $\mathcal{T}_0$ of the oscillations at rest.

In summary, we have considered decay laws at rest which reproduce an oscillating decay rate, Eq. (\ref{sqrtP0toy1}). We have described the oscillating decay laws in the laboratory frame, via Eqs. (\ref{PptHYJosc})-(\ref{Phi}), the exponential-like regime of the oscillating decay laws, via Eq. (\ref{sqrtPptoy1}) and Eqs. (\ref{PPExpOscapprox})-(\ref{PpOscapprox}), the time window over which the oscillating decay laws are exponentially damped, via Eq. (\ref{ExpTExp0}), and the transformed modal decay width
and frequency. Under special conditions the survival probability and the period of the damped oscillations transforms according to the relativistic dilation of times, Eq. (\ref{PpPprel}) and Eq. (\ref{TpT0}). These descriptions and properties constitute the first central result of the paper.

\begin{figure*}
 \includegraphics[width=0.5\textwidth]{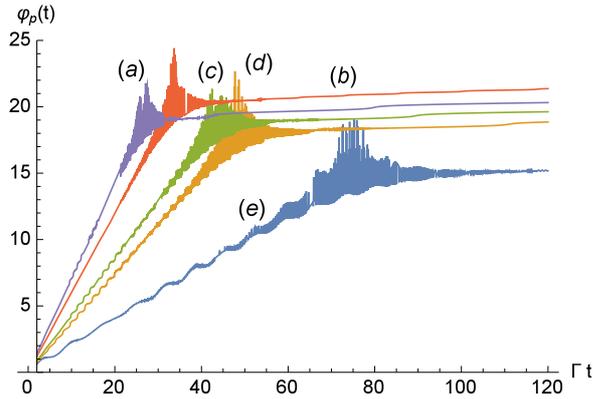}
\caption{(Color online) Function $\varphi_p(t)$ versus $ \Gamma t$, for $ 2 \leq \Gamma t  \leq 120$, and different values of the ratios $p/ \Gamma$ and $M/\Gamma$ and of the corresponding Lorentz factor $\gamma$. The survival probability at rest $\mathcal{P}_0(t)$ is described via Eq. (\ref{sqrtP0toy1}) with different values of the frequency $\Omega$ and of the amplitude $a$. Curve $(a)$ corresponds to $p/ \Gamma=100$, $M/\Gamma=100$, $\gamma=\sqrt{2}$, $\Omega/\Gamma=10$ and $a=0.04$. Curve $(b)$ corresponds to $p/ \Gamma=200$, $M/\Gamma=150$, $\gamma\simeq 1.6667$, $\Omega/\Gamma=40$ and $a=0.01$. Curve $(c)$ corresponds to $p/ \Gamma=210$, $M/\Gamma=100$, $\gamma\simeq 2.3259$, $\Omega/\Gamma=10$ and $a=0.04$. Curve $(d)$ corresponds to $p/ \Gamma=200$, $M/\Gamma=80$, $\gamma\simeq 2.6926$, $\Omega/\Gamma=10$ and $a=0.04$. Curve $(e)$ corresponds to $p/ \Gamma=150$, $M/\Gamma=30$, $\gamma\simeq 2.6926$, $\Omega/\Gamma=5$ and $a=0.09$.}
\label{fig3}
\end{figure*}

Numerical analysis of the function $\varphi_p(t)$ is displayed in Figure \ref{fig3} in case the survival probability is described by the oscillating decay law (\ref{sqrtP0toy1}). The linear growth of the function $\varphi_p(t)$, which is predicted theoretically over the time window (\ref{ExpTExp0}) by the scaling laws (\ref{PPExpOscapprox})-(\ref{PpOscapprox}), is confirmed numerically by the approximate linear behaviors which are observed in Figure \ref{fig3}.

\section{Transformation of decay laws with more exponentially damped oscillating modes}\label{4}

The analysis which is performed in the previous Section is generalized in the present Section to decay laws which are composed by more oscillating modes. In fact, the modulus of the survival amplitude at rest is approximated by the superposition of an arbitrary finite number of purely exponential modes and exponentially damped oscillating modes with decay widths $\Gamma_1, \ldots$, $\Gamma_N$, and frequencies of oscillations $\Omega_1, \ldots$, $\Omega_N$,
\begin{eqnarray}
\sqrt{\mathcal{P}_0(t)}=\sum_{j=1}^N w_j 
\exp\left(- \frac{\Gamma_j}{2}t\right) \left(1-a_j+a_j \cos\left(\Omega_j t\right)\right),
\label{sqrtP0toyjN}
\end{eqnarray}
where $0<\Gamma_1<\ldots<\Gamma_N$. The constraints $w_j>0$, $0<a_j<1/2$, $\Omega_j<M$, $\Gamma_j/\left(M-\Omega_j\right) \ll 1$, $2 a \Omega_j/\sqrt{1-2a_j}<\Gamma_j$ are required to hold for every $j=1,\ldots,N$. The weights are normalized to unity, $\sum_{j=1}^N w_j =1$. In this way, the survival probability is canonically decreasing, $\dot{\mathcal{P}}_0(t)<0$ for every $t>0$, and the canonical value $\mathcal{P}_0(0)=1$ is recovered, in addition to the asymptotic limit $\mathcal{P}_0\left( \infty\right)=0$. 
The survival probability at rest $\mathcal{P}_0(t)$, which is obtained from Eq. (\ref{sqrtP0toyjN}), is the sum of the exponential-like term $\mathcal{P}^{\rm (exp)}_0(t)$ and of the damped oscillating term $\mathcal{P}^{\rm (osc)}_0(t)$,
\begin{eqnarray}
&&\hspace{-4em}\mathcal{P}_0(t)=\mathcal{P}^{\rm (exp)}_0\left(t\right)
+\mathcal{P}^{\rm (osc)}_0(t), \label{P0Nt}
\end{eqnarray}
where
\begin{eqnarray}
&&\hspace{-4em}\mathcal{P}^{\rm (exp)}_0(t)=
\sum_{j=1}^{N}\sum_{l=1}^{N}
w_j w_l \left(\left(1-a_j\right)\left(1-a_l\right)+\frac{\upsilon^{\prime}_{j,l}}{2}a_j a_l\right) 
\exp\left(-\frac{\Gamma_j+\Gamma_l}{2} t\right), \label{P0expN}
\\
&&\hspace{-4em}\mathcal{P}^{\rm (osc)}_0(t)=
\sum_{j=1}^{N}\sum_{l=1}^{N} w_j w_l a_j 
\exp\left(-\frac{\Gamma_j+\Gamma_l}{2} t\right)
\Bigg(2 \left(1-a_l\right) \cos \left(\Omega_j t\right)
\nonumber \\
&&\hspace{1.0em}+\frac{a_l}{2}\Bigg(\cos \left(\left(\Omega_j+\Omega_l\right) t\right)+  \upsilon^{\prime \prime}_{j,l}   \cos \left(\left(\Omega_j-\Omega_l\right) t\right)\Bigg)\Bigg).
\label{PpOsc12Napprox}
\end{eqnarray}
The coefficient $\upsilon^{\prime}_{j,l}$ is unity for $j=l$, and for $j \neq l$ if $\Omega_j=\Omega_l$, and vanishes otherwise. The coefficient $\upsilon^{\prime\prime}_{j,l}$ vanishes for $j=l$, and for $j \neq l$ if $\Omega_j=\Omega_l$, and is unity otherwise. The decay rate at rest results to be the following product,
\begin{eqnarray}
\hspace{-2em}\dot{\mathcal{P}}_0(t)=-\sqrt{\mathcal{P}_0(t)}
\sum_{j=1}^N w_j\exp\left(-\frac{\Gamma_j}{2}t\right)\left(\lambda_{1,j}+\lambda_{2,j} 
\cos \left(\Omega_j t-\beta_j\right)\right), \label{D1P0toyj}
\end{eqnarray}
 which consists in a superposition of purely exponential modes and exponentially damped oscillating modes with more frequencies of oscillations. The involved coefficients are defined by the expressions below,
\begin{eqnarray}
\hspace{-2em}\lambda_{1,j}=\Gamma_j\left(1-a_j\right),\hspace{1em} \lambda_{2,j}=a_j\sqrt{\Gamma_j^2+4\Omega_j^2},\hspace{1em} \beta_j = \arccos \frac{\Gamma_j}{\sqrt{\Gamma_j^2+4 \Omega_j^2}},\nonumber
\end{eqnarray}
for every $j=1,\ldots,N$. Again, we stress that the survival probability at rest can not be approximated by Eq. (\ref{sqrtP0toyjN}) over very short and very long times. In fact, according to Eq. (\ref{D1P0toyj}), the decay rate is negative, $\dot{\mathcal{P}}_0(0) = -\sum_{j=1}^N w_j \Gamma_j$, instead of vanishing, for $t=0$.

If the modulus of the survival amplitude at rest is given by Eq. (\ref{sqrtP0toyjN}) with the required conditions, the survival probability $\mathcal{P}_p(t)$ is approximated in the laboratory reference frame $\mathfrak{S}_p$ by the expression below,
\begin{eqnarray}
&&\hspace{-4em}\mathcal{P}_p(t)\simeq\Bigg|
 \sum_{j=1}^N w_j\left(K\left(M,\Gamma_j,p,\Omega_j,a_j,t\right)
+\imath \frac{p \Gamma_j}{\pi M^2} \Phi\left(M,p,\Omega_j,a_j,t\right)\right)\Bigg|^2, \label{PptHYJoscN}
\end{eqnarray}
over times $t$ such that either $t>1/\left(10 \Gamma_j\right)$ or $\left(M-\Omega_{j}\right) t \gg 1$, for every $j=1,\ldots,N$. 
For the sake of simplicity, expression (\ref{PptHYJoscN}) certainly approximates the transformed survival probability $\mathcal{P}_p(t)$ over times $t$ such that either $t>1/\left(10 \Gamma_1\right)$ or $\left(M-\Omega_{ \rm max}\right) t \gg 1$. The frequency $\Omega_{ \rm max}$ is the maximum among the frequencies $\Omega_1, \ldots,\Omega_N$.

\subsection{Time window for decay laws with more exponentially damped oscillating modes in the laboratory reference frame}\label{41}

At this stage, we estimate the exponential times in the laboratory reference frame $\mathfrak{S}_p$. We refer to these times, if they exist, as the times over which the modulus of the transformed survival amplitude results in superpositions of purely exponential modes and exponentially damped oscillating modes. For the sake of convenience, we introduce the shifted masses $M^+_1, \ldots,M^+_N$ and $M^-_1, \ldots,M^-_N$, the shifted decay widths $\Gamma^+_1, \ldots, \Gamma^+_N$ and $\Gamma^-_1,  \ldots, \Gamma^-_N$, and the shifted relativistic Lorentz factors $\gamma^+_1,  \ldots, \gamma^+_N$
and $\gamma^-_1,  \ldots, \gamma^-_N$, via the following expressions,
\begin{eqnarray}
M^{\mp}_j=M\mp \Omega_j,\hspace{1em}\Gamma^{\mp}_j= \frac{\gamma}{\gamma^{\mp}_j} \Gamma_j,\hspace{1em}
\gamma^{\mp}_j=\sqrt{1+\frac{p^2}{\left(M_j^{\mp}\right)^2}}, \nonumber
\end{eqnarray}
for every $j=1,\ldots,N$. 
Exponentially damped oscillations are provided to the expression (\ref{PptHYJoscN}) of the survival probability by the functions $K\left(M,\Gamma_1,p,\Omega_1,a_1,t\right), \ldots$, $K\left(M,\Gamma_N,p,\Omega_N,a_N,t\right)$. See appendix for the description of the exponentially damped oscillations of the transformed survival probability $\mathcal{P}_p(t)$ and for details.

The transformation of the oscillating decay laws is simplified in case the modal frequencies of oscillations are small compared to the mass of resonance, $\Omega_{\rm max} \ll M$. In the laboratory reference frame $\mathfrak{S}_p$ the exponential times are estimated by studying the order of magnitude of the parameter $\xi_j^{\prime}$ which is defined as below,
 \begin{eqnarray}
\xi_j^{\prime}=\sqrt{\frac{\Gamma_j }{\pi  M}\sqrt{1-\frac{1}
{\gamma^2}}}\sum_{l=1}^N\frac{ w_l \Gamma_l W\left(M,\Omega_l,a_l\right)}{2 M w_j\left(1-2 a_j\right)} 
, \label{xijosc}
\end{eqnarray}
for every $j=1,\ldots,N$. The indexes $j_1,\ldots,j_{n_1}$ are chosen among the indexes $1,\ldots,N$, in such a way that $\xi^{\prime}_{j_l} \ll 10^{-2}$ for every $l=1,\ldots,n_1$. Notice that this constraint is not fulfilled in the limiting case $a_{j_l} \to 1/2^-$. The chosen indexes are sorted in increasing order, $j_1<\ldots<j_{n_1}$, with $1\leq n_1 \leq N$. The set $I^{'}_p$ of the exponential times is given by the following union,
\begin{eqnarray}
I^{'}_p=\bigcup_{l=1}^{n_1} I_{p,l}, \label{Ipprime}
\end{eqnarray}
where \begin{eqnarray}
I_{p,l}=\left[\frac{2\zeta_{\rm min}\gamma}{\Gamma_{j_l}},\frac{2\zeta_{\rm max}\gamma}{\Gamma_{j_l}}\right], \label{IpI0l}
\end{eqnarray}
for every $l=1,\ldots,n_1$. The set $I^{'}_p$ is represented by one single closed interval if $n_1=1$ and in other cases. If $n_1 >1$, let the condition 
\begin{eqnarray}
\frac{\Gamma_{j_l}}{\Gamma_{j_{l+1}}}> \frac{\zeta_{\rm min}}{\zeta_{\rm max}}\simeq 1.83 \times 10^{-5}, \label{cond1windows}
\end{eqnarray}
hold for every $l=1,\ldots,n_1-1$. Then, in the laboratory reference frame $\mathfrak{S}_p$, the set $I^{'}_p$ coincides with the closed interval $\left[2 \zeta_{\rm min} \gamma/\Gamma_{j_{n_1}},2 \zeta_{\rm max} \gamma/\Gamma_{j_{1}}\right]$. Consequently, the following time window,
\begin{eqnarray}
\frac{2 \zeta_{\rm max} }{\Gamma_{j_1}}\gamma\gtrsim t\gtrsim\frac{2 \zeta_{\rm min} }{\Gamma_{j_{n_1}}}\gamma, \label{ExpTn0n1}
\end{eqnarray}
estimates the exponential times in the laboratory reference frame $\mathfrak{S}_p$, in case $\Omega_{\rm max} \ll M$. The present approach requires that the functions $\Phi\left(M,p,\Omega_1,a_1,t\right), \ldots, \Phi\left(M,p,\Omega_N,a_N,t\right)$ provide the inverse-power-law behavior $1/\sqrt{pt}$ to the expression (\ref{PptHYJoscN}) of the transformed survival probability over the exponential times. Consequently, over the set $I^{'}_p$ of the exponential times the following constraints must hold, either $t>1/\left(10 \Gamma_{j_l}\right)$ or $M^-_{j_l} t \gg 1$, for every $l=1,\ldots,n_1$, and $p t \gg 1$. These constraints are certainly fulfilled if either $t>1/\left(10 \Gamma_{1}\right)$ or $\left(M-\Omega_{\rm max}\right) t \gg 1$, and $p t \gg 1$. The constraint $t>1/\left(10 \Gamma_1\right)$ is fulfilled over the set $I^{'}_p$ if $20 \zeta_{\rm min}\gamma>\Gamma_{j_{n_1}}/\Gamma_1$. If this constraint is not realized, the condition $\left(M-\Omega_{\rm max}\right) t \gg 1$ holds over the set $I^{'}_p$ if $2 \zeta_{\rm min}\gamma\gg\Gamma_{j_{n_1}}/\left(M-\Omega_{ \rm max}\right)$. The condition $p t \gg 1$ holds over the set $I^{'}_p$ if $2 \zeta_{\rm min}\gamma \sqrt{\gamma^2-1}\gg\Gamma_{j_{n_0}}/M$. The nonrelativistic limit, $\gamma\to 1^+$, is excluded. Generally, the above constraints are fulfilled except for strong decays and in the nonrelativistic regime. If the above constraints do not hold, the method which is reported in the previous paragraph can be adjusted by choosing a different order of magnitude for the parameter $\xi_j$ and by changing the values of the parameters $\zeta_{\rm min}$ and $\zeta_{\rm max}$ accordingly. See Ref. \cite{Gxiv2019} for details.

The transformed survival probability $\mathcal{P}_p(t)$ is approximated by the form
\begin{eqnarray}
\hspace{-0em}\mathcal{P}_p(t)\simeq \left|\sum_{l} {}^{'} 
w_{j_l}\exp\left(- \frac{\Gamma_{j_l}}{2 \gamma}t\right)
\left(1-a_{j_l}+ a_{j_l}\cos\left( \frac{\Omega_{j_l}}{\gamma} t\right)\right)
\right|^2,
\label{PptoyjNOmegasmall}
\end{eqnarray}
over the set $I^{'}_p$ of the exponential times, or over the time window (\ref{ExpTn0n1}) if $n_1=1$ or if $n_1>1$ and condition (\ref{cond1windows}) is fulfilled for every $l=1, \ldots, n_1-1$. The sum $\sum_{l} {}^{'}$ depends on time as follows. For every instant $t$ which belongs to the set $I^{'}_p$ of the exponential times, the sum is performed over every index $l$ which is selected among the indexes $1,\ldots,n_1$, in such a way that $t \in I_{p,l}$. Expression (\ref{PptoyjNOmegasmall}) of the survival probability $\mathcal{P}_p(t)$ results to be the sum, Eq. (\ref{PPExpOscapprox}), of the exponential-like term $\mathcal{P}^{\rm (exp)}_p(t)$ and of the damped oscillating term $\mathcal{P}^{\rm (osc)}_p(t)$, which are given by the following expressions,
\begin{eqnarray}
&&\hspace{-5em}\mathcal{P}^{\rm (exp)}_p(t)\simeq 
\frac{1}{2}
\sum_{l} {}^{'}\sum_{l^{'}} {}^{'} \upsilon^{'}_{j_l,j_{l^{'}}}w_{j_l}w_{j_{l^{'}}} a_{j_l}a_{j_{l^{'}}} \exp\left(- \frac{\Gamma_{j_l}+\Gamma_{j_{l^{'}}}}{2 \gamma}t\right) \nonumber \\ &&\hspace{0.2em}+\left(\sum_{l} {}^{'} w_{j_l} \left(1-a_{j_l}\right)\exp\left(- \frac{\Gamma_{j_l}}{2 \gamma}t\right)\right)^2, \label{PptoyjNOmegasmallExp} 
\\
&&\hspace{-5em}\mathcal{P}^{\rm (osc)}_p(t)\simeq 
2\sum_{l} {}^{'}\sum_{l^{'}} {}^{'}w_{j_l}w_{j_{l^{'}}}a_{j_{l^{'}}}\left(1-a_{j_l}\right)\exp\left(- \frac{\Gamma_{j_l}+\Gamma_{j_{l^{'}}}}{2 \gamma}t\right)
\cos\left(\frac{\Omega_{j_{l^{'}}}}{\gamma}t\right)\nonumber \\
&&\hspace{0.0em}+
\frac{1}{2}\sum_{l} {}^{'}\sum_{l^{'}} {}^{'} w_{j_l}w_{j_{l^{'}}} a_{j_l}a_{j_{l^{'}}} \exp\left(- \frac{\Gamma_{j_l}+\Gamma_{j_{l^{'}}}}{2 \gamma}t\right) 
\nonumber \\
&&\hspace{0.0em} \times 
\left(
\cos\left(\frac{\Omega_{j_{l}}+\Omega_{j_{l^{'}}}}{\gamma}t\right)
+\upsilon^{''}_{j_l,j_{l^{'}}}\cos\left(\frac{\Omega_{j_{l}}-\Omega_{j_{l^{'}}}}{\gamma}t\right)\right).
\label{PptoyjNOmegasmallOsc}
\end{eqnarray}

In the rest reference frame $\mathfrak{S}_0$ consider the set of times
$I^{'}_0$ which is defined by the following union,
\begin{eqnarray}
I^{'}_0=\bigcup_{l=1}^{n_1} I_{0,l}, \label{I0primeprime}
\end{eqnarray}
where
\begin{eqnarray}
I_{0,l}=\left[\frac{2\zeta_{\rm min}}{\Gamma_{j_l}},\frac{2\zeta_{\rm max}}{\Gamma_{j_l}}\right],
\label{I0l}
\end{eqnarray}
for every $l=1,\ldots,n_1$. By substituting the ratio $t/\gamma$ with $t$ in Eq. (\ref{PptoyjNOmegasmall}), the technique which is described in the second paragraph of the present subsection selects the dominant modes of the modulus of the survival amplitude at rest over the set $I^{'}_0$. Refer to \cite{Gxiv2019} for details. Consequently, Eq. (\ref{PPExpOscapprox}) and the scaling laws (\ref{PpPprel}), (\ref{PpExpapprox}), (\ref{PpOscapprox}) hold over the set $I^{'}_p$ of the exponential times, in the laboratory reference frame $\mathfrak{S}_p$. In the rest reference frame $\mathfrak{S}_0$ the scaling laws holds over the set $I^{'}_0$ of times. If $n_1=1$, or if $n_1 >1$ and condition (\ref{cond1windows}) is fulfilled for every $l=1, \ldots, n_1-1$, the set $I^{'}_0$ coincides with the closed interval $\left[2 \zeta_{\rm min} /\Gamma_{j_{n_1}},2 \zeta_{\rm max} /\Gamma_{j_{1}}\right]$ and represents the time windows,
\begin{eqnarray}
\frac{2 \zeta_{\rm max} }{\Gamma_{j_1}}\gtrsim t\gtrsim\frac{2 \zeta_{\rm min} }{\Gamma_{j_{n_1}}}, \label{ExpTn0prime}
\end{eqnarray}
in the rest reference frame $\mathfrak{S}_0$. Consider the set $I^{'}_p$ of the exponential times in the laboratory reference frame $\mathfrak{S}_p$. Over these times the transformed decay laws are composed by transformed purely exponential modes and transformed exponentially damped oscillating modes. The modal decay width at rest $\Gamma_{j_l}$ transforms in the laboratory reference frame in the reduced decay width $\Gamma_{j_l}/\gamma$, and the modal frequency of oscillation at rest $\Omega_{j_l}$ transforms in the reduced frequency $\Omega_{j_l}/\gamma$, for every $l=1,\ldots,n_1$.

Let the interval $ \left[t_{p,0},t_{p,1}\right]$ exist in the set $I^{'}_p$ of the exponential times such that the modes which compose the sum $\sum_{l} {}^{'}$ do not change for every time $t \in \left[t_{p,0},t_{p,1}\right]$. Let the maximum modal frequency involved be multiple of the remaining modal frequencies of oscillations. This means that relation $\Omega_{j_l}=\Omega^{'}_{\rm max}/k_l $, where $k_l$ is a nonvanshing natural number, holds for every index $l$ over which the sum $\sum_{l} {}^{'}$ is performed. The frequency $\Omega^{'}_{\rm max}$ is the maximum among the frequencies $\Omega_{j_l}$ of the exponentially damped oscillating modes which are selected for the evaluation of the sum $\sum_{l} {}^{'}$ over the interval $\left[t_{p,0},t_{p,1}\right]$. Under the above-mentioned conditions, expression (\ref{PptoyjNOmegasmall}) of the survival probability is approximately periodic over the interval $\left[t_{p,0},t_{p,1}\right]$ in the laboratory reference frame $\mathfrak{S}_p$. The period $\mathcal{T}^{'}_p$ of these oscillations is 
\begin{eqnarray}
\mathcal{T}^{'}_p\simeq \gamma\frac{2 \pi  }{\Omega^{'}_{\rm max}}, \label{TpprimeN}
\end{eqnarray}
and diverges in the ultrarelativistic limit, $p \gg M$.
 Similarly, the damped oscillations of the survival probability at rest $\mathcal{P}_0(t)$ are approximately periodic over the interval $\left[ t_{0,0}, t_{0,1}\right]$, 
where $t_{0,0}=t_{p,0}/\gamma $ and $t_{0,1}=t_{p,1}/\gamma$. The period $\mathcal{T}^{'}_0$ of these oscillations reads 
\begin{eqnarray}
\mathcal{T}^{'}_0\simeq \frac{2 \pi  }{\Omega^{'}_{\rm max}}.
\label{T0primeN}
\end{eqnarray}
Relations (\ref{TpprimeN}) are (\ref{TpprimeN}) are examined below by considering the relativistic dilation of times.

\subsection{Transformation of times and relativistic time dilation in decay laws with more exponentially damped oscillating modes}\label{42}

At this stage, we describe the transformation of times by evaluating the function $\varphi_p(t)$ which is defined in Section \ref{32} via Eqs. (\ref{PpP0def}) and (\ref{phipP0Pp}). The survival probability at rest $\mathcal{P}_0(t)$ is given by Eq. (\ref {sqrtP0toyjN}) and is a canonically decreasing function of time for the selected values of the involved parameters. In general, the function $\mathcal{P}^{-1}_0\left(r\right)$ is evaluated numerically for the oscillating decay laws under study. Once the function $\mathcal{P}^{-1}_0\left(r\right)$ is obtained, the function $\varphi_p(t)$ is estimated by the following expression, 
\begin{eqnarray}
&& \hspace{-5em}\varphi_p(t)\simeq \mathcal{P}^{-1}_0\Bigg(\Bigg|
 \sum_{j=1}^N w_j\left(K\left(M,\Gamma_j,p,\Omega_j,a_j,t\right)
+\imath \frac{p \Gamma_j}{\pi M^2} \Phi\left(M,p,\Omega_j,a_j,t\right)\right)\Bigg|^2\Bigg), \label{phipP0PpHYJoscN}
\end{eqnarray} 
 over times $t$ such that either $M^-_j t \gg 1$ or $t>1/\left(10 \Gamma_j\right)$, for every $j=1,\ldots,N$.

 If $\Omega_{\rm max} \ll M$ the survival probability transforms, approximately, according the scaling law (\ref{PpPprel}) over the set $I^{'}_p$ of the exponential times in the laboratory reference frame $\mathfrak{S}_p$. This scaling law reproduces the relativistic dilation of times in case the mass of resonance M is considered to be the rest mass of the moving unstable system. The function $\varphi_p(t)$ exhibits the approximately linear growth (\ref{varphilinear}) over the set $I^{'}_p$ of the exponential times, or over the time window (\ref{ExpTn0n1}) if $n_1=1$ or if $n_1>1$ and the constraint (\ref{cond1windows}) is fulfilled for every $l=1, \ldots, n_1-1$. The constraints which are described in the second paragraph of Section \ref{41} are required to hold over the exponential times. 
Under the above-reported conditions, the transformed survival probability is approximately periodic in the interval $\left[t_{p,0},t_{p,1}\right]$. Approximately periodic oscillations of the survival probability at rest appear in the rest reference frame over the interval $\left[t_{0,0},t_{0,1}\right]$. The period of the transformed oscillations $\mathcal{T}^{'}_p$, given by Eq. (\ref{TpprimeN}), is related to the period of the approximately periodic oscillations at rest $\mathcal{T}^{'}_0$, given by Eq. (\ref{T0primeN}), by the following relation,
\begin{eqnarray}
\mathcal{T}^{'}_p\simeq \gamma\mathcal{T}^{'}_0. \label{TpT0primeN}
\end{eqnarray}
This relation represents the relativistic time dilation of the period $\mathcal{T}^{'}_0$ of the damped oscillations at rest if the mass of resonance $M$ is considered to be the mass at rest of the unstable quantum system which moves with constant linear momentum $p$. Notice that the 
duration of the periodic oscillations, $\left(t_{p,1}-t_{p,0}\right)$, in the laboratory reference frame $\mathfrak{S}_p$ is the relativistic time dilation of the periodic oscillations at rest, $\left(t_{0,1}-t_{0,0}\right)$, in the rest reference frame $\mathfrak{S}_0$.

In summary, we have considered oscillating decay laws at rest which consist in superpositions of an arbitrary finite number of purely exponential modes and exponentially damped oscillating modes with different frequencies of oscillations, Eq. (\ref{sqrtP0toyjN}). We have described the oscillating decay laws in the laboratory reference frame, via Eq. (\ref{PptHYJoscN}), the exponential-like regime of the oscillating decay laws, via Eqs. (\ref{PptoyjNOmegasmall})-(\ref{PptoyjNOmegasmallOsc}), the exponential times, via Eqs. (\ref{xijosc})-(\ref{ExpTn0n1}), and the decay widths and the frequencies of oscillations of the transformed modes. Under special conditions, the purely exponential term, Eq. (\ref{P0expN}), and the exponentially damped oscillating term, Eq. (\ref{PpOsc12Napprox}), of the survival probability transform according to the same scaling law, Eqs. (\ref{PpExpapprox}) and (\ref{PpOscapprox}) and Eqs. (\ref{PptoyjNOmegasmallExp}) and (\ref{PptoyjNOmegasmallOsc}), respectively. Under determined conditions the transformed survival probability exhibits damped periodic oscillations. If the mass of resonance is interpreted as the mass at rest of the moving unstable quantum system, the survival probability and the period of the oscillations, transform, approximately, according to the relativistic dilation of times, Eq. (\ref{PpPprel}) and Eq. (\ref{TpT0primeN}), respectively. These descriptions and properties constitute the the last of the main results of the paper.

\section{Summary and conclusions}\label{5}

The appearance of oscillations in the decay rate of unstable systems is a peculiar phenomenon which has attracted a great deal of attention. This interest follows, especially, the detection in the GSI experiment of oscillations which are superimposed on the canonical exponential decay laws \cite{GSI1,GSI2}. This kind of oscillations are obtained, theoretically, in the decay laws of unstable quantum system by introducing deviations in the Breit-Wigner form of the MDD \cite{GPoscillatingDecaysQM2012,GPoscillatingDecaysQM2012PoS}. 
Usually, unstable systems move in the laboratory reference frame where the decay laws are detected. The general transformation of the decay laws at rest, which is induced by the change of reference frame, is described via quantum theory and special relativity \cite{HEP_Stef1996,HEP_Shir2004,HEP_Shir2006,GiacosaAPPB2016,GiacosaAPPB2017,UrbAPB2017,GiacosaAHEP2018}. The transformed survival amplitude consists in an integral form which involves the model-independent MDD and the linear momentum of the moving unstable quantum system. This integral form was adopted to study the transformation of oscillating decay laws over short and long times in case the decaying system is initially prepared in superpositions of two, approximately orthogonal, unstable quantum states \cite{HEP_Shir2004,GJPA2018}. The corresponding MDDs are bounded from below and are represented by truncated Breit-Wigner forms \cite{HEP_Shir2004} or exhibit thresholds in the (different) non-vanishing lower bounds of the mass spectra \cite{GJPA2018}. The transformation of the period of the oscillations is determined by the features of the MDDs and by the linear momentum of the moving unstable quantum system.

As a continuation of the above-mentioned scenario, in the present research work we have considered general forms of the modulus of the survival amplitude at rest which consist in superpositions of purely exponential and exponentially damped oscillating modes. These forms provide general expressions of the survival probability at rest which decay monotonically and exhibit oscillating decay rates. These expressions approximate the oscillating decay laws at rest of unstable quantum systems over intermediate times. We have studied the transformation of the oscillating decay laws at rest which is induced by the change of reference frame. The transformed decay laws and the transformed times have been determined in the laboratory reference frame where the unstable quantum system moves with constant linear momentum, by assuming that the MDD is approximately symmetric with respect to the mass of resonance. By considering the modal frequencies of oscillations to be small compared to the mass of resonance, time intervals are determined over which the transformed decay laws consist in the superposition of transformed purely exponential modes and transformed exponentially damped oscillating modes. The modal decay widths at rest, $\Gamma_j$, transform regularly in reduced decay widths, $\Gamma_j/\gamma$, and the modal frequencies at rest, $\Omega_j$, transform regularly in reduced frequencies, $\Omega_j/\gamma$, in the laboratory reference frame. Equivalently, over the selected times, both the purely exponential modes and the exponentially damped oscillating modes at rest transform, independently and regularly, according to the same time scaling. The time intervals constitute one single time window if the modal decay widths fulfill determined conditions. If the oscillations of the decay laws at rest are approximately periodic over the time window, the transformed decay laws are approximately periodic in the laboratory reference frame over the transformed time window and the period of the oscillations transforms, regularly, according to the time scaling. The relativistic dilation of times is reproduced by the time scaling, over the time window, if the mass of resonance of the MDD is considered to be the mass at rest of the moving unstable quantum system with relativistic Lorentz factor $\gamma$. By adopting this interpretation, the survival probability at rest, the duration of the time window in the rest reference frame, the period of the damped oscillations at rest, if the oscillations are periodic, and the modal frequencies of the oscillations transform according to the relativistic dilation of times, by changing reference frame.

In conclusion, the decay laws of moving unstable systems with oscillating decay rates exhibit, over determined intermediate times, some regularities in the transformation which is induced by the change of reference frame. Interpreting the experimental works on oscillating decay rates of moving unstable systems is beyond the purposes of the present paper. However, we believe that further insight about the description of the oscillating decay laws via quantum theory and special relativity may be provided by decomposing the detected oscillating decay laws into the above-mentioned purely exponential and exponentially damped oscillating modes and searching for the above-mentioned regularities.

\appendix\label{A}
\section{Details}

The transformation of the oscillating decay law at rest which is described by Eq. (\ref{sqrtP0toy1}) is obtained from Eq. (\ref{PpP0Int1}). The involved integrals are studied in Ref \cite{HEP_Shir2004}. In this way, Eqs. (\ref{PptHYJosc})-(\ref{Xi}) are determined. The function $K\left(M,\Gamma,p,\Omega,a,t\right)$, which appears in the form (\ref{PptHYJosc}) of the transformed survival probability, is properly approximated by the following sum of exponential terms,
\begin{eqnarray}
&&\hspace{-4.5em}K\left(M,\Gamma,p,\Omega,a,t\right)\simeq 
\frac{a}{2}\exp \left(-\frac{t}{2}\left(\frac{\Gamma^-}{\gamma}+ 2\imath M^- \gamma^-\right)\right)+
\left(1-a\right)\nonumber \\ &&\hspace{4.9em}\times \exp \left(-\frac{t}{2}\left(\frac{\Gamma}{\gamma}+ 2\imath M \gamma\right)\right)+\frac{a}{2}
\exp \left(-\frac{t}{2}\left(\frac{\Gamma^+}{\gamma}+ 2\imath M^+ \gamma^+\right)\right), 
\label{PiExpOsc11}
\end{eqnarray}
as the condition $\Gamma/M^- \ll 1$ holds. Instead, the function $\Phi\left(M,\Gamma,p,\Omega,a,t\right)$ contributes to the form (\ref{PptHYJosc}) with an inverse power law over times $t$ such that $pt \gg 1$,
\begin{eqnarray}
\hspace{-2em} \imath \frac{p \Gamma}{\pi M^2} \Phi
\left(M,p,\Omega,a,t\right)\simeq 
\frac{\exp\left(\imath\left(pt-3\pi/4\right)\right)}{\sqrt{2 \pi p t}}
\frac{p \Gamma}{M^2}W\left(M,\Omega,a\right).
\label{IPLTermOsc}
\end{eqnarray} According to Eq. (\ref{PiExpOsc11}), the function $K\left(M,\Gamma,p,\Omega,a,t\right)$ provides to the expression (\ref{PptHYJosc}) purely exponential decays with transformed decay widths $\Gamma^{-}/\gamma$, $\Gamma/\gamma$, $\Gamma^{+}/\gamma$, and exponentially damped oscillations with frequencies $\left|M^- \gamma^--M \gamma\right|$, $\left|M^+ \gamma^+-M \gamma\right|$, $\left|M^- \gamma^--M^+ \gamma^+\right|$.

The relations which are reported below help to estimate the exponential times. The following inequalities,
\begin{eqnarray}
\hspace{-2em}\frac{1}{2}<\sqrt{\frac{1+p^2/M^2}{1+4 p^2/\left( M^2\right)}}<\frac{\gamma}{\gamma_-}<1<\frac{\gamma}{\gamma_+}<\sqrt{\frac{1+p^2/M^2}{1+4 p^2/\left(9 M^2\right)}}<\frac{3}{2}, \label{ineqgpm12} 
\end{eqnarray}
hold for $\Omega<M/2$, $0<a<1/2$ and for every value of the linear momentum $p$. 
The following approximations of the transformed decay widths,
\begin{eqnarray}
&&\hspace{0em}\frac{\Gamma^{\mp}}{\gamma}=\frac{\Gamma}{\gamma^{\mp}}\simeq \frac{\Gamma}{\gamma}\left(1\mp\frac{p^2\Omega}{\gamma^2 M^3 }\right), \label{ineqGammagamma12}
\end{eqnarray}
and of the transformed frequencies of the oscillations,
\begin{eqnarray}
&&\hspace{-5em}M^{\mp} \gamma^{\mp}-M \gamma\simeq\mp \frac{\Omega}{\gamma}\left(1 \mp \frac{p^2\Omega}{2 \gamma^2 M^3}\right),
\hspace{1em} M^+ \gamma^+-M^- \gamma^-\simeq \frac{\Omega}{\gamma}\left(2-\frac{p^2 \Omega^2}{M^4 \gamma^4}\right), \label{ineqMgpmOmega12} 
\end{eqnarray}
hold for $\Omega \ll M$. The function $W\left(M,\Omega,a\right)$, which appears in Eq. (\ref{xiprime}), is defined as below,
\begin{eqnarray}
W\left(M,\Omega,a\right)=1+a \frac{\Omega^2}{M^2}\frac{3-\Omega^2/M^2}{\left(1-\Omega^2/M^2\right)^2},\label{Wdef}
\end{eqnarray}
for every value of the mass of resonance $M$ and for every allowed value of the frequency $\Omega$ and of the parameter $a$. The inequality 
\begin{eqnarray}
1<W\left(M,\Omega,a\right)<\frac{29}{18}, \label{ineqW1}
\end{eqnarray}
holds for $\Omega<M/2$, $0<a<1/2$ and for every value of the mass of resonance $M$.

In case $\Omega\ll M$, the function $K\left(M,\Gamma,p,\Omega,a,t\right)$ is approximated by the form below,
\begin{eqnarray}
&&\hspace{-5em}K\left(M,\Gamma,p,\Omega,a,t\right)
\simeq \exp \left(-\frac{t}{2}\left(\frac{\Gamma}{\gamma}+2\imath M \gamma \right)\right)\left(1-a+ a \cos\left(\frac{\Omega}{\gamma} t\right)\right)
.\label{PiapproxCos1}
\end{eqnarray}
This expression suggests that Eq. (\ref{sqrtPptoy1}) properly approximates the transformed decay law over the time window (\ref{ExpTExp0}) in the laboratory reference frame $\mathfrak{S}_p$. These times are obtained by requiring the following condition,
\begin{eqnarray}
&&\hspace{0.0em}\left(1-2a\right)\exp \left(-\frac{\Gamma t}{2\gamma}\right) \gg
\frac{p \Gamma}{\pi M^2} \left|\Phi
\left(M,p,\Omega,a,t\right)\right|,
\label{ExpToscPiggPhi}
\end{eqnarray}
to hold for $p t \gg 1$. The above relation is studied by analyzing the order of magnitude of the parameter $\xi^{\prime}$ which is given by Eq. (\ref{xiprime}). See Ref. \cite{Gxiv2019} for details. The comparison between Eq. (\ref{sqrtP0toy1}) and Eq. (\ref{sqrtPptoy1}) leads to Eqs. (\ref{PPExpOscapprox})-(\ref{TpT0}).

The transformation of the oscillating decay law at rest which is described by Eq. (\ref{sqrtP0toyjN}) is obtained from Eq. (\ref{PpP0Int1}) and results in Eq. (\ref{PptHYJoscN}). The transformed purely exponential modes and the transformed exponentially damped oscillating modes are provided by the functions $K\left(M,\Gamma_1,p,\Omega_1,a_1,t\right), \ldots$, $K\left(M,\Gamma_N,p,\Omega_N,a_N,t\right)$, as the condition $\Gamma_j/M^-_j \ll 1$ holds for every $j=1,\ldots,N$, 
\begin{eqnarray}
&&\hspace{-5em}\sum_{j=1}^N w_j K\left(M,\Gamma_j,p,\Omega_j,a_j,t\right)
\simeq 
\sum_{j=1}^N
w_j\Bigg(\frac{a_j}{2}\exp \Bigg(-\frac{t}{2}\Bigg(\frac{\Gamma^{-}_j}{\gamma}+ 2\imath M^{-}_{j} \gamma^{-}_{j}\Bigg)\Bigg)\nonumber \\ &&\hspace{-5em}
+\left(1-a_j\right)\exp \Bigg(-\frac{t}{2}\Bigg(\frac{\Gamma_j}{\gamma}+ 2\imath M \gamma\Bigg)\Bigg)
+\frac{a_j}{2}
\exp \Bigg(-\frac{t}{2}\Bigg(\frac{\Gamma_j^{+}}{\gamma}+ 2\imath M^{+}_{j} \gamma^{+}_{j}\Bigg)\Bigg)
\Bigg).\label{ExpTermOscN}
\end{eqnarray}
Instead, the functions $\Phi\left(M,\Gamma_1,p,\Omega_1,a_1,t\right), \ldots$, $\Phi\left(M,\Gamma_N,p,\Omega_N,a_N,t\right)$ contribute to the expression (\ref{PptHYJoscN}) of the survival probability with the inverse power law $1/\sqrt{pt}$ for $pt \gg1$,
\begin{eqnarray}
\hspace{-3em} \imath\sum_{j=1}^N w_j \frac{p \Gamma_j}{\pi M^2} \Phi
\left(M,p,\Omega_j,a_j,t\right)\simeq 
\frac{\exp\left(\imath\left(pt-3\pi/4\right)\right)}{\sqrt{2 \pi p t}}
\sum_{j=1}^N w_j\frac{p \Gamma_j}{M^2}W\left(M,\Omega_j,a_j\right).
\label{IPLTermOscN}
\end{eqnarray} 
 Differently form the decay widths $\Gamma_1, \ldots,\Gamma_N$, the frequencies $\Omega_1, \ldots,\Omega_N$ are not sorted in increasing or decreasing order and, in general, can even coincide. Therefore, some simplifications are performed by introducing the parameters $w^{\prime}_1, \ldots, w^{\prime}_{N^{\prime}}$, $\Gamma^{\prime}_1, \ldots, \Gamma^{\prime}_{N^{\prime}}$ and $M^{\prime}_1, \ldots, M^{\prime}_{N^{\prime}}$. These parameters and the natural number $N^{\prime}$ are defined via the following equality, 
\begin{eqnarray}
&&\hspace{-3.6em}\sum_{j=1}^N
w_j\Bigg(\frac{a_j}{2}\exp \Bigg(-\frac{t}{2}\Bigg(\frac{\Gamma^{-}_j}{\gamma}+ 2\imath M^{-}_{j} \gamma^{-}_{j}\Bigg)\Bigg)+\left(1-a_j\right)\exp \Bigg(-\frac{t}{2}\Bigg(\frac{\Gamma_j}{\gamma}+ 2\imath M \gamma\Bigg)\Bigg)
\nonumber \\ &&\hspace{-3.6em}
+\frac{a_j}{2}
\exp \Bigg(-\frac{t}{2}\Bigg(\frac{\Gamma_j^{+}}{\gamma}+ 2\imath M^{+}_{j} \gamma^{+}_{j}\Bigg)\Bigg)
\Bigg)=\sum_{i=1}^{N^{\prime}}w^{\prime}_j \exp
\Bigg(-\frac{t}{2}\Bigg(\frac{\Gamma^{\prime}_j}{\gamma}+ 2\imath M^{\prime}_j \gamma\Bigg)\Bigg),
 \label{changeExpOsc12}
\end{eqnarray}
by sorting the decay widths in increasing order, $\Gamma^{\prime}_1< \ldots<\Gamma^{\prime}_{N^{\prime}}$. Therefore, the transformed decay widths are determined by simplifying and ordering the decay widths $\Gamma^{-}_1,\Gamma_1, \Gamma^{+}_1,\ldots$, $\Gamma^{-}_N,\Gamma_N, \Gamma^{+}_N$. The right side of Eq. (\ref{changeExpOsc12}) contributes to the expression (\ref{PptHYJoscN}) of the survival probability with purely exponential modes and exponentially damped oscillations modes. The dominant purely exponential mode $\exp\left(-\Gamma^{\prime}_1 t/\left(2 \gamma\right)\right)$ and the fastest purely exponential mode $\exp\left(-\Gamma^{\prime}_{N^{\prime}} t/\left(2 \gamma\right)\right)$ are determined by the following decay widths,
\begin{eqnarray}
\hspace{-2em}\Gamma^{\prime}_1=\min \left\{\Gamma^-_j, \,\,\,\forall\,\,j=1,\ldots,N\right\}, \hspace{1em} \Gamma^{\prime}_{N^{\prime}}=\max \left\{\Gamma^+_j, \,\,\,\forall\,\,j=1,\ldots,N\right\}. \nonumber
\end{eqnarray}
The weights $w^{\prime}_1, \ldots$, $w^{\prime}_{N^{\prime}}$ are normalized to unity, $\sum_{i=1}^{N^{\prime}}w^{\prime}_j=1$. The selected frequencies of the oscillations are chosen among the non-vanishing values of the form $\gamma\left|M^{\prime}_l-M^{\prime}_{l^{\prime}}\right|$, which appears in the right side of Eq. (\ref{changeExpOsc12}).

The exponential times are obtained by requiring that the condition 
\begin{eqnarray}
w_{j_l} \left(1-2 a_{j_l}\right)\exp\left(\frac{-\Gamma_{j_l}t}{2 \gamma}\right)
\gg \frac{1}{\sqrt{2 \pi p t}}
\sum_{i=1}^N w_i\frac{p \Gamma_i}{M^2}W\left(M,\Omega_i,a_i\right),
\label{ExpToscPiggPhiNt}
\end{eqnarray}
holds over times $t$ such that either $t>1/\left(10 \Gamma_1\right)$ or $\left(M-\Omega_{\rm max}\right) \gg 1$, and $p t \gg 1$. In this way, the $j_l$th term of the sum $\sum_{l} {}^{'}$, which appear in Eq. (\ref{PptoyjNOmegasmall}), dominates the inverse power law (\ref{IPLTermOscN}) over the set $I^{'}_{p,l}$ of the exponential times. The parameter $\xi^{\prime}_j$ is defined via Eq. (\ref{xijosc}) by following Ref. \cite{Gxiv2019}. Relation (\ref{ExpToscPiggPhiNt}) is equivalent to the inequality $\xi^{\prime}_{j_l}\ll 10^{-2}$ and holds over the time interval $I_{p,l}$ which is given by Eq. (\ref{IpI0l}). The set $I^{'}_p$ is defined via Eq. (\ref{Ipprime}) and the exponential time window is approximated via Eq. (\ref{ExpTn0n1}) in case $n_1=1$, or $n_1>1$ and condition (\ref{cond1windows}) is hold for every $l=1,\ldots,n_1-1$. The comparison between Eq. (\ref{PptoyjNOmegasmall}) and Eq. (\ref{sqrtP0toyjN}) suggests the scaling properties of the survival probability, of the purely exponential term $\mathcal{P}^{\rm (exp)}_p(t)$ and of the oscillating term $\mathcal{P}^{\rm (osc)}_p(t)$. The form (\ref{phipP0PpHYJoscN}) of the function $\varphi_p(t)$ is obtained from expression (\ref{PptHYJoscN}) of the survival probability $\mathcal{P}_p(t)$. If the survival probability is approximated by Eq. (\ref{PptoyjNOmegasmall}), the scaling law (\ref{PpPprel}) holds and the linear growth (\ref{varphilinear}) appears over the set $I^{'}_p$ of the exponential times. This concludes the demonstration of the present results.

\end{document}